\documentclass[aps,pra]{revtex4}
\usepackage{graphicx}

\begin{document}

\title{Network for transfer of an arbitrary $n$-qubit atomic state via cavity QED
\thanks{Email: zhangzj@wipm.ac.cn}}

\author{Zhan-jun Zhang$^{1,2}$\\
{\normalsize $^1$ School of Physics \& Material Science, Anhui University, Hefei 230039, China} \\
{\normalsize $^2$ Wuhan Institute of Physics and Mathematics,
Chinese Academy of Sciences, Wuhan 430071, China} \\
{\normalsize Email: zhangzj@wipm.ac.cn}}

\date{\today}
\maketitle

\begin{minipage}{420pt}
I show a scheme which allows a perfect transfer of an unknown
single-qubit atomic state from one atom to another by letting two
atoms interact simultaneously with a cavity QED. During the
interaction between atom and cavity, the cavity is only virtually
excited and accordingly the scheme is insensitive to the cavity
field states and cavity decay. Based on this scheme, a network for
transfer of an arbitrary single-qubit atomic state between atoms
is engineered. Then the scheme is generalized to perfectly
transfer an arbitrary 2-qubit atomic state and accordingly a
network for transfer of an arbitrary 2-qubit atomic state is
designed. At last, it is proven that the schemes can be
generalized to an arbitrary $n(n\ge 3)$-qubit atomic state
transfer case and a corresponding network is also proposed. \\

\end{minipage}

{\bf I. Introduction}\\

No cloning theorem forbids a perfect copy of an arbitrary unknown
quantum state. How to interchange different resources has ever
been a question in quantum computation and quantum information. In
1993, Bennett {\it et al}[1] first presented a quantum
teleportation scheme. In the scheme, an arbitrary unknown quantum
state in a qubit can be teleported to a distant qubit with the the
help of Einstein-Podolsky-Rosen (EPR) pair. Their work showed in
essence the interchangeability of different resources in quantum
mechanics. Hence, after Bennett {\it et al}'s pioneering work in
1993, quantum teleportation has attracted many attentions[2-27].
However, in the experimental aspects, since the complete Bell
state set can not be discriminated so far, teleportation can only
be achieved in a nondeterministic manner. As a matter of fact, one
can transfer quantum states ont only by the method of
teleportation[1] but also by quantum state transfer
network[28-32]. The basic idea of a quantum network is to transfer
a quantum state from one node to another with aid of a {\it
messenger} such that it arrives intact. In this process, the
unknown state initially in a node is first transferred to the
messenger, then the messenger carrying the unknown state goes
ahead to the destination node. When it arrives, the unknown state
in the messenger is further transferred to the destination node.
For different networks, messengers may be different. For examples,
in Ref.[28-30], the messenger is a photon, which delivers the
unknown state from one atom to another, while in Ref.[32] the
messenger is an atom, which delivers an unknown state from a
cavity to another. In fact, recently quantum network has attracted
some attentions and interests continuously increase[28-35]. There
are various physical systems can that can serve as quantum
networks, one of them being the atom-cavity system[28-32]. In this
paper I will propose a scheme for transfer of an unknown state
from one atom to another via direct interaction with each other in
a cavity. I will focus on the situation in which state transfer is
perfect. Hence in the designed network the unknown state can be
transferred in a deterministic way to another atom which is
elsewhere. The distinct advantage of the present scheme is that
during the passage of the atoms through the cavity field, the
cavity is only virtually excited. No transfer of quantum
information will occur between the atoms and cavity. The present
scheme does not require any measurement on atoms.

The present paper is organized as follows. In section 2, I
describe the scheme for a perfect transfer of an unknown
single-qubit atomic state from one atom to another by letting two
atoms interact simultaneously with a cavity QED. As a result of
the scheme, a network for transfer of an arbitrary single-qubit
atomic state between atoms is engineered. In section 3,the scheme
is generalized to perfectly transfer an arbitrary 2-qubit atomic
state and accordingly a network for transfer of an arbitrary
2-qubit atomic state is designed. A prove of an arbitrary $n(n\ge
3)$-qubit atomic state can be perfectly  transferred in a similar
network is shown in the section 4. A brief summary is given in
the last section.\\

{\bf II. Network for transfer of an arbitrary single-qubit atomic
state via cavity QED}\\

Assume that the atom 1 is in a state,
\begin{eqnarray}
|\psi\rangle_1=\alpha|0\rangle_1+ \beta|1\rangle_1,
\end{eqnarray}
where $\alpha$ and $\beta$ are {\it unknown} arbitrary
coefficients, $|\alpha|^2+|\beta|^2=1$. $|1\rangle$ and $
|0\rangle$ are the excited and ground states of the atom,
respectively. This arbitrary state in the atom 1 needs to be
transferred to another atom which is elsewhere.

Before giving the present state transfer scheme it is helpful to
briefly review the teleportation scheme proposed by Ye and
Guo[23]. The Ye-Guo scheme includes the following steps (cf.,
figure 1(a)): (i) Two identical two-level atoms 2 and 3 are
prepared in the state $|1\rangle_2$ and $|0\rangle_3$,
respectively. Each of the two identical cavities $C_1$ and $C_2$
is prepared in a vacuum state. (ii) Let atoms 2 and 3 interact
simultaneously with the cavity $C_1$. By the way, in the case
$\Delta\gg g$, no energy exchange between the atomic system and
the cavity exists, where $\Delta$ is the detuning between the
atomic transition frequency $\omega_0$ and the cavity frequency
$\omega$, and $g$ is the atom-cavity coupling constant. In the
interaction picture, the effective Hamiltonian for this system is
[36]
\begin{eqnarray}
H_e=\lambda (\sum\limits_{k=1}^2|1\rangle_{kk}\langle 1|
+\sum\limits_{k,j=1,k\neq j}^2 S_k^+S_j^-),
\end{eqnarray}
where $\lambda=g^2/\Delta$, $S_k^+=|1\rangle_{kk}\langle 0|$,
$S_k^-=|0\rangle_{kk}\langle 1|$, and $|1\rangle_k$ and $
|0\rangle_k$ are the excited and ground states of the $k$th atom,
respectively. For the following different initial atomic states,
they evolve as follows,
\begin{eqnarray}
|1\rangle_k|0\rangle_j &\rightarrow& e^{-i\lambda t}(\cos\lambda t
|1\rangle_k|0\rangle_j -i\sin\lambda t |0\rangle_k|1\rangle_j, \\
|0\rangle_k|1\rangle_j &\rightarrow& e^{-i\lambda t}(\cos\lambda t
|0\rangle_k|1\rangle_j -i\sin\lambda t |1\rangle_k|0\rangle_j, \\
|1\rangle_k|1\rangle_j &\rightarrow& e^{-2i\lambda t}
|1\rangle_k|1\rangle_j, \\
|0\rangle_k|0\rangle_j &\rightarrow& |0\rangle_k|0\rangle_j,
\end{eqnarray}
where $t$ is the interaction time. With the choice of $\lambda
t=\pi/4$, the initial state of the two-atom (atoms 2 and 3) system
evolves to a maximally entangled state,
\begin{eqnarray}
|\psi\rangle_{23}=\frac{e^{-i\pi/4}}{\sqrt{2}}(
|1\rangle_2|0\rangle_3-i |0\rangle_2|1\rangle_3) .
\end{eqnarray}
Note that in the equations 4-7 of Ref.[23] an unimportant overall
factor $e^{-i\pi/4}$ is missed and a factor $\frac{1}{\sqrt{2}}$
should be included in the last term of the equation 7. (iii) Let
atoms 1 and 2 interact simultaneously with the cavity $C_2$ with
the interaction time such that $\lambda t=\pi/4$, then the state
of the whole atomic system becomes
\begin{eqnarray}
|\psi\rangle_{123}= \frac{-i}{\sqrt 2}&\{&\frac{1}{\sqrt
2}|0\rangle_1|1\rangle_2(\alpha|0\rangle_3-
\beta|1\rangle_3)+e^{-i\pi/4}\beta
|1\rangle_1|1\rangle_2|0\rangle_3 \nonumber \\ &+& \frac{1}{\sqrt
2}|1\rangle_1|0\rangle_2(\alpha|0\rangle_3+
\beta|1\rangle_3)+e^{-i\pi/4}\alpha
|0\rangle_1|0\rangle_2|1\rangle_3\}.
\end{eqnarray}
(iv) If the states of the atoms 1 and 2 are detected to be
$|1\rangle_1$ and $|0\rangle_2$ respectively after they pass
through the cavity $C_2$, then the state of atom 3 is exactly the
unknown state initially in atom 1 after its passage through the
cavity $C_1$. If the state of the atoms 1 and 2 is
$|0\rangle_1|1\rangle_2$, then atom 3 needs to perform a phase
transformation to recover the initial state of atom 1. If the
atoms 1 and 3 are detected in the state $|1\rangle_1|1\rangle_2$
or $|0\rangle_1|0\rangle_2$, the teleportation fails. Hence, in
the Ye-Guo scheme [23], the teleportation is {\it
nondeterministic} and its success probability is $1/2$.

\begin{figure}
\begin{center}\vskip -4cm
\includegraphics[width=1.0\textwidth]{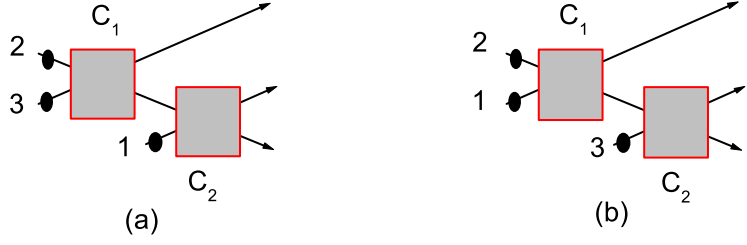}
\vskip -5cm \caption{ The black dots represent atoms. The grey box
stands for cavity. The initial states of the atoms 2 and 1 are
respectively $|1\rangle_2$ and  $\alpha|0\rangle_1+
\beta|1\rangle_1$  before they pass through the cavities $C_1$ and
$C_2$. (a) The initial state of the atom 3 is $|0\rangle_3$. The
interaction time of the two atoms interact simultaneously with
each cavity satisfies $\lambda t=\pi/4$. (b) The initial state of
the atom 3 is $|1\rangle_3$. The interaction time of the two atoms
interact simultaneously with each cavity satisfies $\lambda
t=\pi/2$. See text for detail.} \label{f1}
\end{center}
\end{figure}

Now let us move to present my scheme. In my scheme almost the same
device as that in Ye-Guo's scheme are employed. However, it allows
a {\it perfect} transfer of the unknown state in atom 1 to atom 3.
One will see this later. My scheme includes the following steps
(cf., figure 1(b)): (i) This step is same as that in the Ye-Guo
scheme except that the two-level atom 3 is prepared in the state
$|1\rangle_3$ instead of $|0\rangle_3$. (ii) Let the atoms 1 and 2
instead of the atoms 2 and 3 in the Ye-Guo scheme interact
simultaneously with the cavity 1. The initial state of the atoms 1
and 2 before their passage through the cavity $C_1$ is
\begin{eqnarray}
|\psi\rangle_{12}=(\alpha|0\rangle_1+
\beta|1\rangle_1)|1\rangle_2.
\end{eqnarray}
After an interaction time $t$, the whole system evolves to
\begin{eqnarray}
|\psi'\rangle_{12}=\alpha e^{-i\lambda t} (\cos\lambda t
|0\rangle_1|1\rangle_2-i\sin\lambda t|1\rangle_1|0\rangle_2)+
\beta e^{-2i\lambda t}|1\rangle_1 |1\rangle_2.
\end{eqnarray}
Different from the interaction time in the Ye-Guo schme, the
interaction time in the present scheme is doubled, that is,
$\lambda t=\pi/2$ is chosen. Then the final state of the atoms 1
and 2 after the evolution is
\begin{eqnarray}
|\psi''\rangle_{12}=-|1\rangle_1(\alpha|0\rangle_2+
\beta|1\rangle_2).
\end{eqnarray}
Obviously, the unknown state initially in the atom 1  is
transferred to the atom 2 after they interact in the cavity $C_1$
with the certain time. (iii) Similarly, if atoms 2 and 3
simultaneously interact with the cavity $C_2$ with the interaction
time satisfying $\lambda t=\pi/2$, then the final state of the
total system is
\begin{eqnarray}
|\psi\rangle_{123}=|1\rangle_1|1\rangle_2(\alpha|0\rangle_3+
\beta|1\rangle_3).
\end{eqnarray}
One can easily find that, after the interactions, the unknown
state initially in the atom 1 is further transferred to the atom
3, and the atoms 1, 2 and 3 do not entangle with each other. Hence
the measurements on the atoms 1 and 2, which need to be performed
in the Ye-Guo scheme to collapse the entire state and accordingly
to get or reconstruct the unknown state, are unnecessary in the
present scheme. If the similar procedures are repeated, then the
unknown state can be transferred to an atom which is elsewhere
(cf., figure 2). Thus far, a network for transfer of an arbitrary
single-qubit atomic state between atoms is established.

Compared to the Ye-Guo teleportation scheme[23], the distinct
advantages of the present scheme are: (1) The present scheme
provides a {\it deterministic} way to transfer the unknown state
between atoms via cavity QED; (2) Measurements on other atoms are
unnecessary; (3) After the whole transfer process all atoms are
still in their initially prepared states (i.e., the excited
states) except for the first and last atoms, hence they can be
reused for next state transfer in the network. (4) Each cavity
acts in essence as a SWAP gate except for an overall unimportant
factor. Moreover, as same as stressed by Ye and Guo in Ref.[23],
during the interaction between atom and cavity, the cavity is only
virtually excited and thus the present scheme is insensitive to
the cavity field states and cavity decay.
This is also an advantage of the present scheme.\\

\begin{figure}
\begin{center}\vskip -4cm
\includegraphics[width=1.0\textwidth]{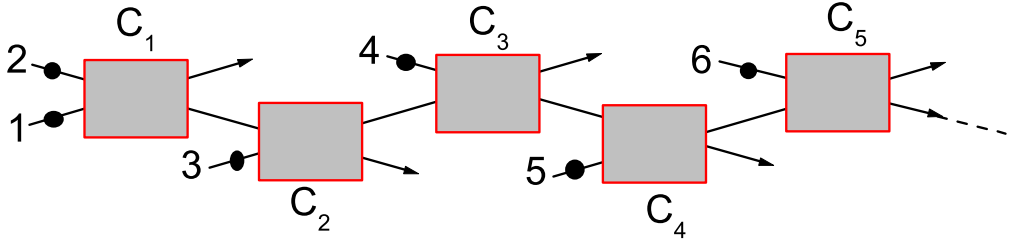}
\vskip -6cm \caption{The black dots represent atoms. The grey box
stands for cavity. The initial states of the atom 1  is
$\alpha|0\rangle+ \beta|1\rangle$ and other are in $|1\rangle$'s
before they pass through cavities. The interaction time of each
two atoms interact simultaneously with the corresponding cavity
satisfies $\lambda t=\pi/2$. See text for detail.} \label{f2}
 \end{center}
\end{figure}

{\bf III. Network for transfer of an arbitrary 2-qubit atomic
state via cavity QED} \\

Now let us generalize the above scheme to an arbitrary two-qubit
atomic state transfer scheme. A two-qubit atomic state which needs
to be transferred is written as
\begin{eqnarray}
|\psi\rangle_{a_1a_2}=\alpha|0\rangle_{a_1}|0\rangle_{a_2}+
\beta|0\rangle_{a_1}|1\rangle_{a_2}+\gamma|1\rangle_{a_1}|0\rangle_{a_2}+
\delta|1\rangle_{a_1}|1\rangle_{a_2},
\end{eqnarray}
where $a_1$ and $a_2$ label two atoms, $\alpha$,$\beta$,$\gamma$
and $\delta$ are {\it unknown} complex coefficients, and I assume
$|\psi\rangle_{a_1a_2}$ to be normalized. This generalized scheme
contains the following steps (cf., figure 3): (i) Two atoms $b_1$
and $b_2$ are prepared in excited states, i.e., $|1\rangle_{b_1}$
and $|1\rangle_{b_2}$. Cavities $C_{11}$ and $C_{21}$ are in
vacuum modes. Then the initial state of the system including atoms
$a_1$,$a_2$,$b_1$ and $b_2$ is
\begin{eqnarray}
|\psi\rangle_{a_1a_2b_1b_2}=(\alpha|0\rangle_{a_1}|0\rangle_{a_2}+
\beta|0\rangle_{a_1}|1\rangle_{a_2}+\gamma|1\rangle_{a_1}|0\rangle_{a_2}+
\delta|1\rangle_{a_1}|1\rangle_{a_2})|1\rangle_{b_1}|1\rangle_{b_2}.
\end{eqnarray}
It can be rewritten as
\begin{eqnarray}
|\psi\rangle_{a_1a_2b_1b_2}=(\alpha|0\rangle_{a_1}+\gamma|1\rangle_{a_1})
|1\rangle_{b_1}|0\rangle_{a_2}|1\rangle_{b_2}+
(\beta|0\rangle_{a_1}+\delta|1\rangle_{a_1})|1\rangle_{b_1}|1\rangle_{a_2}|1\rangle_{b_2}.
\end{eqnarray}
(ii) Let atoms $a_1$ and $b_1$ interact simultaneously with the
cavity $C_{11}$ with an interaction time $t$ such that $\lambda
t=\pi/2$, then according to the scheme in the last section, the
state of the system evolves to
\begin{eqnarray}
|\psi'\rangle_{a_1a_2b_1b_2}=-|1\rangle_{a_1}(\alpha|0\rangle_{b_1}+\gamma|1\rangle_{b_1})
|0\rangle_{a_2}|1\rangle_{b_2}-
|1\rangle_{a_1}(\beta|0\rangle_{b_1}+\delta|1\rangle_{b_1})|1\rangle_{a_2}|1\rangle_{b_2}.
\end{eqnarray}
It can be rewritten as
\begin{eqnarray}
|\psi'\rangle_{a_1a_2b_1b_2}=-|1\rangle_{a_1}|0\rangle_{b_1}(\alpha|0\rangle_{a_2}
+\beta|1\rangle_{a_2})|1\rangle_{b_2}
-|1\rangle_{a_1}|1\rangle_{b_1}(\gamma|0\rangle_{a_2}
+\delta|1\rangle_{a_2})|1\rangle_{b_2} .
\end{eqnarray}
(iii) Similarly, let atoms $a_2$ and $b_2$ interact simultaneously
with the cavity $C_{21}$ with an interaction time $t$ such that
$\lambda t=\pi/2$, then the state of the system evolves to
\begin{eqnarray}
|\psi'\rangle_{a_1a_2b_1b_2}&=&|1\rangle_{a_1}|1\rangle_{a_2}|0\rangle_{b_1}(\alpha|0\rangle_{b_2}
+\beta|1\rangle_{b_2})+
|1\rangle_{a_1}|1\rangle_{a_2}|1\rangle_{b_1}(\gamma|0\rangle_{b_2}
+\delta|1\rangle_{b_2}) \nonumber \\
&=&
|1\rangle_{a_1}|1\rangle_{a_2}(\alpha|0\rangle_{b_1}|0\rangle_{b_2}
+\beta|0\rangle_{b_1}|1\rangle_{b_2}+\gamma|1\rangle_{b_1}|0\rangle_{b_2}
+\delta|1\rangle_{b_1}|1\rangle_{b_2}) \nonumber \\
&=&|1\rangle_{a_1}|1\rangle_{a_2}|\psi\rangle_{b_1b_2} .
\end{eqnarray}
Obviously, the unknown two-qubit atomic state initially in the
atoms $a_1$ and $a_2$ is transferred to the atoms $b_1$ and $b_2$,
and vice versa. Hence the two cavities $C_{11}$ and $C_{21}$ act
as a SWAP gate of two-qubit atomic states. (iii) If similar
procedures are repeated (cf., figure 3), then the arbitrary
two-qubit atomic state can be further transferred to $c_1$ and
$c_2$, $d_1$ and $d_2$, and so on (cf., Figure 3). Thus a network
for transfer of an arbitrary 2-qubit atomic state between atoms is
established.

\begin{figure}
\begin{center}\vskip -3cm
\includegraphics[width=1.0\textwidth]{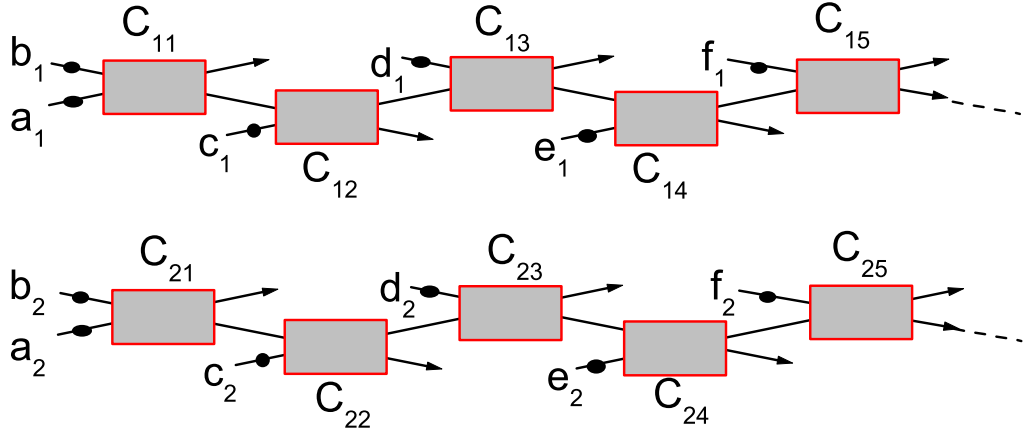}
\vskip -4cm \caption{The network for transfer of an arbitrary
2-qubit state. See text for detail.} \label{f3}
 \end{center}
\end{figure}

One can easily find that the advantages mentioned in last section
remain for this generalized scheme. Moreover, in Ref.[23], Ye and
Guo have proposed a scheme for teleportion of a {\it special}
class of unknown two-qubit atomic states by using a GHZ atomic
state. The successful teleportation can only be achieved with a
probability of $1/2$. In contrast, in the present scheme the
transfer of an {\it arbitrary} two-qubit atomic states but not a
special class of two-qubit entangled atom states can be achieved
{\it deterministically} and no complicated initial states need to
be prepared. These are additional advantages of this generalized scheme. \\

{\bf IV. Network for transfer of an arbitrary $n(n\ge 3)$-qubit
atomic state via cavity QED} \\

Now let us generalize the above scheme to an arbitrary $n(n\ge
3)$-qubit atomic state transfer scheme. In the following I will
prove that, if an arbitrary $(n-1)(n\ge 3)$-qubit quantum state
can be transferred successfully by using $n-1$ cavities in a
network, then an arbitrary $n(n\ge 3)$-qubit quantum state can be
transferred successfully via using $n$ cavities. An arbitrary
$n(n\ge 3)$-qubit atomic state which needs to be transferred is
written as
\begin{eqnarray}
|\xi\rangle_{a_1a_2\dots a_n}= \sum\limits_{m_n=0}^1\dots
\sum\limits_{m_2=0}^1\sum\limits_{m_1=0}^1C_{m_1m_2\dots
m_n}|m_1\rangle_{a_1}|m_2\rangle_{a_2}\dots|m_n\rangle_{a_n},
\end{eqnarray}
where $C$'s are complex coefficients and $|\xi\rangle_{a_1a_2\dots
a_n}$ is assumed to be normalized. It can be decomposed as
\begin{eqnarray}
|\xi\rangle_{a_1a_2\dots a_n}&=&
|0\rangle_{a_n}(\sum\limits_{m_{n-1}=0}^1\dots
\sum\limits_{m_2=0}^1\sum\limits_{m_1=0}^1C_{m_1m_2\dots
m_{n-1}0}|m_1\rangle_{a_1}|m_2\rangle_{a_2}\dots|m_{n-1}\rangle_{a_{n-1}})
\nonumber \\ &+& |1\rangle_{a_n}(\sum\limits_{m_{n-1}=0}^1\dots
\sum\limits_{m_2=0}^1\sum\limits_{m_1=0}^1C_{m_1m_2\dots
m_{n-1}1}|m_1\rangle_{a_1}|m_2\rangle_{a_2}\dots|m_{n-1}\rangle_{a_{n-1}})\nonumber\\
&\equiv& |0\rangle_{a_n}\zeta_{a_1a_2\dots
a_{n-1}}+|1\rangle_{a_n}\zeta'_{a_1a_2\dots a_{n-1}}.
\end{eqnarray}
Here $\zeta_{a_1a_2\dots a_{n-1}}$ and $\zeta'_{a_1a_2\dots
a_{n-1}}$ are in essence arbitrary $(n-1)$-qubit states,
respectively.

\begin{figure}
\begin{center}\vskip -3cm
\includegraphics[width=1.0\textwidth]{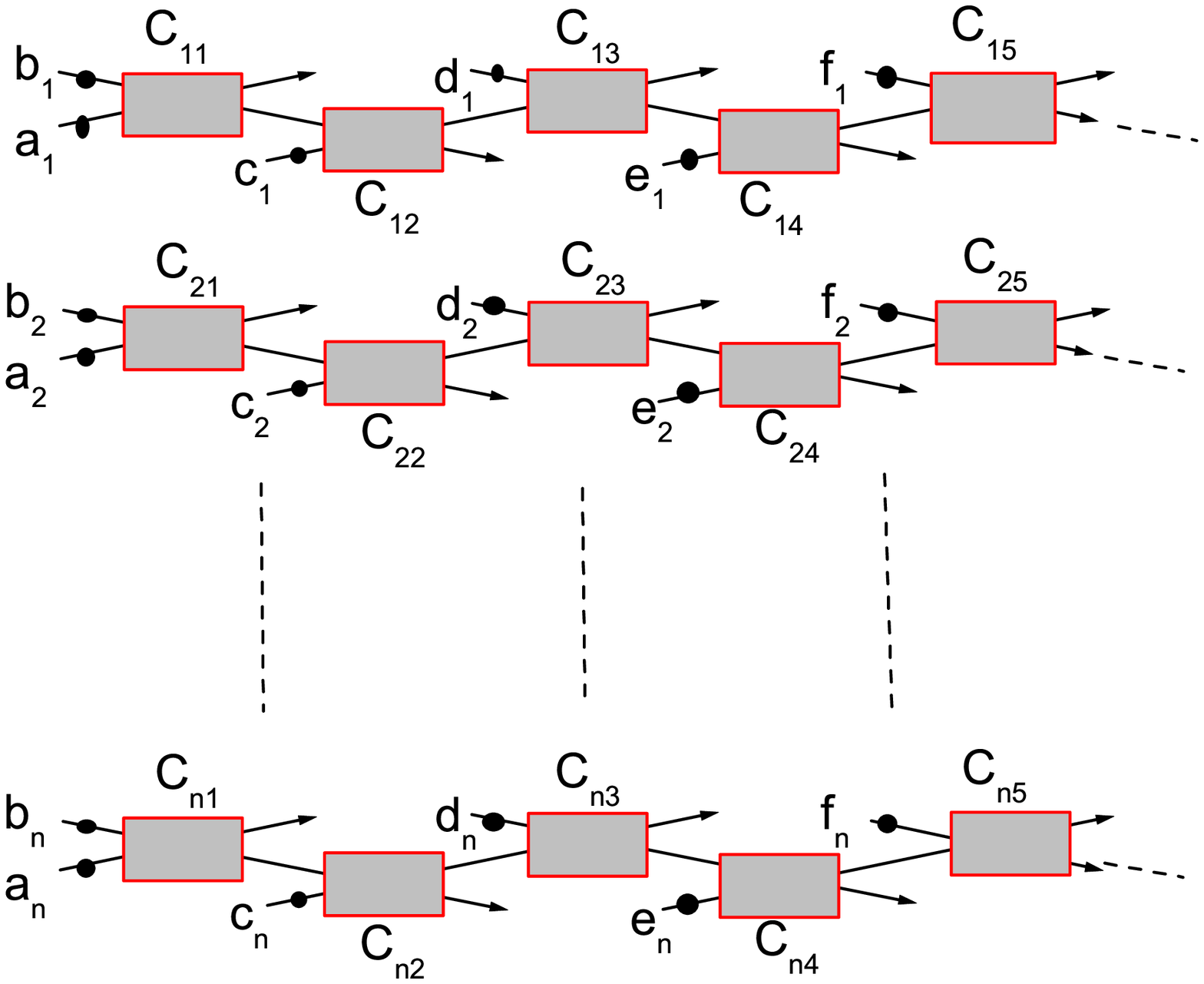}
\vskip -4cm \caption{The network for transfer of an arbitrary
$n$-qubit state. See text for detail.} \label{f4}
 \end{center}
\end{figure}

Suppose there are $n$ atoms labelling as $b_1$, $b_2$, \dots,
$b_n$ and each is prepared in a state $|1\rangle$. Then the
initial joint state of the $2n$ atoms (i.e., $a_1$, $a_2$, \dots,
$a_n$, $b_1$, $b_2$, \dots, $b_n$) is
\begin{eqnarray}
&& |\xi\rangle_{a_1a_2\dots
a_n}|1\rangle_{b_1}|1\rangle_{b_2}\dots
|1\rangle_{b_{n-1}}|1\rangle_{b_n} \nonumber \\&=&
|0\rangle_{a_n}\zeta_{a_1a_2\dots
a_{n-1}}|1\rangle_{b_1}|1\rangle_{b_2}\dots
|1\rangle_{b_{n-1}}|1\rangle_{b_n} +
 |1\rangle_{a_n}\zeta'_{a_1a_2\dots
a_{n-1}}|1\rangle_{b_1}|1\rangle_{b_2}\dots
|1\rangle_{b_{n-1}}|1\rangle_{b_n}.
\end{eqnarray}
It has previously been supposed that an arbitrary $(N-1)(N\ge
3)$-qubit quantum state can be transferred successfully via using
$n-1$ cavities in a network. This means that the above initial
state can evolve to the following state,
\begin{eqnarray}
&& |0\rangle_{a_n}|1\rangle_{a_1}|1\rangle_{a_2}\dots
|1\rangle_{a_{n-1}}\zeta_{b_1b_2\dots b_{n-1}}|1\rangle_{b_n} +
 |1\rangle_{a_n}|1\rangle_{a_1}|1\rangle_{a_2}\dots
|1\rangle_{a_{n-1}}\zeta'_{b_1b_2\dots b_{n-1}}|1\rangle_{b_n}
\nonumber \\ &=& |1\rangle_{a_1}|1\rangle_{a_2}\dots
|1\rangle_{a_{n-1}}(\zeta_{b_1b_2\dots b_{n-1}}|0\rangle_{a_n} +
 \zeta'_{b_1b_2\dots b_{n-1}}|1\rangle_{a_n})|1\rangle_{b_n}.
\end{eqnarray}
Now let the atoms $a_n$ and $b_n$ interact simultaneously with the
$n$th cavity, then  according to equations 3 and 4 the state of
the system evolves as
\begin{eqnarray}
|1\rangle_{a_1}|1\rangle_{a_2}\dots |1\rangle_{a_{n-1}}&[&
\zeta_{b_1b_2\dots b_{n-1}} e^{-i\lambda t} (\cos\lambda t
|0\rangle_{a_n}|1\rangle_{b_n}-i\sin\lambda
t|1\rangle_{a_n}|0\rangle_{b_n})\nonumber \\ && +
\zeta'_{b_1b_2\dots b_{n-1}} e^{-2i\lambda t}|1\rangle_{a_n}
|1\rangle_{b_n}].
\end{eqnarray}
With  a choice of $\lambda t=\pi/2$, then the state is
\begin{eqnarray}
-|1\rangle_{a_1}|1\rangle_{a_2}\dots
|1\rangle_{a_{n-1}}|1\rangle_{a_n}(\zeta_{b_1b_2\dots
b_{n-1}}|0\rangle_{b_n} +
 \zeta'_{b_1b_2\dots b_{n-1}}|1\rangle_{b_n}) =-|1\rangle_{a_1}|1\rangle_{a_2}\dots
|1\rangle_{a_n}|\xi\rangle_{b_1b_2\dots b_n}.
\end{eqnarray}
The equations 19-24 have shown an arbitrary $n(n\ge 3)$-qubit
quantum state can be transferred successfully via using $n$
cavities provided that an arbitrary $(n-1)(n\ge 3)$-qubit quantum
state can be transferred successfully by using $n-1$ cavities in a
network. As a matter of fact, in section 2 I have already shown
that any arbitrary 2-qubit state can be transferred via using two
cavities as a net work. Hence, in terms of recurrence one can
easily conclude that any arbitrary $n(n\ge3)$-qubit state can also
be successfully transferred by using $n$ cavities. According to
this generalized scheme, a network for transfer of an arbitrary
2-qubit atomic state between atoms can be easily established (cf.,
figure
4). \\

{\bf V. Summary} \\

I have explicitly shown a transfer scheme that allows to
faithfully and deterministically transfer an arbitrary $n$-qubit
atomic state to an atom which is elsewhere. The perfect transfer
is realized by letting two atoms interact simultaneously with a
cavity QED. Sicne the large-detuned interaction between driven
atoms and cavity is designed, the cavity is only virtually excited
and accordingly the transfer scheme is insensitive to the cavity
field states and cavity decay. Based on the scheme, a network for
transfer of an arbitrary $n$-qubit atomic state between atoms is
engineered.\\

\noindent {\bf Acknowledgement} \\

This work is supported by the National Natural Science Foundation
of China under Grant No. 10304022. \\

\noindent {\bf References} \\

\noindent[1] C. H. Bennett, G. Brassard C. Crepeau,  R. Jozsa, A.
Peres and W. K. Wotters, Phys. Rev. Lett. {\bf70}, 1895 (1993).

\noindent[2] L. Davidovich {\it et al}, Phys. Rev. {\bf 50}, R895
(1994).

\noindent[3] J. I. Cirac {\it et al}, Phys. Rev. {\bf 50}, R4441
(1994).

\noindent[4] D. Bouwmeester, J. -W. Pan, K. Martle, M. Eibl, H.
Weinfurter, and A. Zeilinger, Nature (London), {\bf 390}, 575
(1997).

\noindent[5] S. B. Zheng and G. C. Guo, Phys. Lett. A {\bf 232},
171 (1997).

\noindent[6] D. Boschi {\it et al}, Phys. Rev. Lett. {\bf 80},
1121 (1998).

\noindent[7] A. Fuusawa, J. L. Sorensen, S. L. Braunstein, C. A.
Fuchs, H. J. Kimble, and E. S. Polzik, Science {\bf 282}, 706
(1998).

\noindent[8] M. A. Nilson, E. Knill, and R. Laflamme, Nature
(London), {\bf 396}, 52 (1998).

\noindent[9] S. Bose, P. L. Knight, M. B. Plenio, and V. Vedral,
Phys. Rev. Lett. {\bf 83}, 5158 (1999).

\noindent[10] M. Ikram, S. Y-. Zhu, and M. S. Zubairy, Phys. Rev.
A {\bf 62}, 022307 (2000).

\noindent[12] S. BAndyopadhyay, Phys. Rev. A {\bf 62}, 012308
(2000).

\noindent[13] W. Son, J. Lee, M. S. Kim, and Y. -J. Park, Phys.
Rev. A {\bf 64}, 064304 (2001).

\noindent[14] J. Lee, H. Min, and S. D. Oh, Phys. Rev. A {\bf 64},
014302 (2001).

\noindent[15] J. Lee, H. Min, and S. D. Oh, Phys. Rev. A {\bf 66},
052318 (2002).

\noindent[16] T. J. Johnson, S. D. Bartlett, and B. C. Sanders,
Phys. Rev. A {\bf 66}, 042326 (2002).

\noindent[17] W. P. Bowen, N. Treps, B. C. Buchler, R. Schnabel,
T. C. Ralph, Hans-A. Bachor, T. Symul, and P. K. Lam, Phys. Rev. A
{\bf 67}, 032302 (2003).

\noindent[18] J. Fang, Y. Lin, S. Zhu, and X. Chen, Phys. Rev. A
{\bf 67}, 014305 (2003).

\noindent[19] N. Ba An, Phys. Rev. A {\bf 68}, 022321 (2003).

\noindent[20] M. Fuji, Phys. Rev. A {\bf 68}, 050302 (2003).

\noindent[21] M. Riebe {\it et al}, Nature (London), {\bf 429},
737 (2004).

\noindent[22] Shi-biao Zheng, Phys. Rev. A {\bf 69}, 064302
(2004).

\noindent[23] Liu Ye and  Guang-can Guo, Phys. Rev. A {\bf 70},
054303 (2004).

\noindent[24] G. Rigolin, Phys. Rev. A {\bf 71}, 032303 (2005).

\noindent[25] Zhan-jun Zhang, Yong Li and Zhong-xiao Man,  Phys.
Rev. A {\bf 71}, 044301 (2005).

\noindent[26] Z. J. Zhang, J. Yang, Z. X. Man and Y. Li, Eur.
Phys. J. D. {\bf 33} 133 (2005).

\noindent[27] Zhan-jun Zhang and Zhong-xiao Man, Phys. Lett. A (In
press).

\noindent[28] J. I. Cirac {\it et al}, Phys. Rev. Lett. {\bf 78},
3221 (1997).

\noindent[29] S. J. van Enk, J. I. Cirac, and P. Zoller, Phys.
Rev. Lett. {\bf 78}, 4293 (1997).

\noindent[30] T. Pellizzari, Phys. Rev. Lett. {\bf 79}, 5242
(1997).

\noindent[31] M. Christandl, N. Datta, A. Ekert, and A. J.
Landahl, Phys. Rev. Lett. {\bf 92}, 187902 (2004).

\noindent[32] A. Asoka and G. S. Agarwal, Phys. Rev. A {\bf 70},
022323 (2004).

\noindent[33] Tao Shi, Ying Li, Zhi Song and Chang-pu Sun, Phys.
Rev. A {\bf 71}, 032309 (2005).

\noindent[34] M. Peternostro, G. M. Palma, M. S. Kim, and G.
Falci, Phys. Rev. A {\bf 71}, 042311 (2005).

\noindent[34] M. H. Yung and S. Bose, Phys. Rev. A {\bf 71},
032310 (2005).

\noindent[35] M. Christandl, N. Datta, T. C. Dorlas, A. Ekert, A.
Kay, and A. J. Landahl, Phys. Rev. A {\bf 71}, 032312 (2005).

\noindent[36] S. B. Zheng and G. C. Guo, Phys. Rev. Lett. {\bf
85}, 2392 (2000).

\enddocument